\documentclass[preprint]{aastex}

\begin{document}

\title {The First Comprehensive Photometric Study of the Algol-type System CL Aurigae}
\author{Jae Woo Lee$^1$, Chun-Hwey Kim$^2$, Duck Hyun Kim$^2$, Seung-Lee Kim$^1$, Chung-Uk Lee$^{1}$, and Robert H. Koch$^3$}
\affil{$^1$Korea Astronomy and Space Science Institute, Daejeon 305-348, Korea}
\email{jwlee@kasi.re.kr, slkim@kasi.re.kr, leecu@kasi.re.kr}
\affil{$^2$Department of Astronomy and Space Science, and Chungbuk National University Observatory, 
		Chungbuk National University, Cheongju 361-763, Korea}
\email{kimch$@$chungbuk.ac.kr, kdh3841@hanmail.net}
\affil{$^3$Department of Physics and Astronomy, University of Pennsylvania, Philadelphia, USA}
\email{rhkoch@earthlink.net}

\begin{abstract}
We present the first extensive photometric results of CL Aur from our $BVRI$ CCD photometry made on 22 nights from 2003 
November through 2005 February. Fifteen new timings of minimum light were obtained. During the past 104 years, 
the orbital period has varied due to a periodic oscillation superposed on a continuous period increase. The period 
and semi-amplitude of the oscillation are about 21.6 yrs and 0.0133 d, respectively.  This detail is interpreted as 
a light-travel-time effect due to a low-luminosity K-type star gravitationally bound to the CL Aur close system.  
Our photometric study indicates that CL Aur is a relatively short-period Algol-type binary with values of $q$=0.602 
and $i$=88$^\circ$.2. Mass transfer from the secondary to the primary eclipsing component is at least partly responsible 
for the observed secular period change with a rate of $dP$/$dt$ = +1.4$\times$10$^{-7}$ d yr$^{-1}$. A cool spot model 
has been calculated but we think that an alternative hot-spot model resulting from a gas stream impact on the hot star 
is more reasonable despite two difficulties with the explanation. Absolute dimensions of the eclipsing system are deduced 
and its present state is compared with tracks for single star and conservative close binary evolution. Finally, we examine 
the possible reconciliation of two different calculations of the luminosity of the hot spot and a re-interpretation of 
the secular term of the period variability.
\end{abstract}

\keywords{binaries: close --- binaries: eclipsing --- stars: individual (CL Aurigae) --- stars: spots}{}

\section{INTRODUCTION}

Algol-type close binaries are semi-detached interacting systems in which one type of interaction is mass transfer 
between the component stars by means of a gas stream. They have been known as good astrophysical laboratories for 
studying accretion processes because a number of them are bright.  They are in the slow phase of mass transfer with 
$dM/dt \simeq 10^{-11}-10^{-7}$ M$_\odot$ yr$^{-1}$ and do not undergo violent eruptions that interfere with 
the accretion process. The circumstellar structures produced by the mass-transfer process in these systems have been 
sorted according to orbital period by Richards \& Albright (1999) but do not depend upon it significantly. Rather, 
their natures can be easily understood from the position of the mass-gaining component in the so-called $r$-$q$ diagram 
in which the fractional radius $r$ = ($R/a$) of a gainer is plotted {\it versus} the mass ratio $q$ and compared with 
the semianalytical computations of the gas stream hydrodynamics of Lubow \& Shu (1975). In the short-period Algols 
located above the $\omega_d$ curve of the diagram (cf. Figure 2 of Richards \& Albright), the hot, detached primary star 
is large relative to the orbital radius and the two components are too close to each other to form an accretion disk or 
even a stable accretion annulus. Instead, it is possible that an impact region, and hence a hot spot, can be formed 
on the surface of the primary star somewhat displaced from the line of centers due to the Coriolis acceleration imposed 
on the flowing gas. If the secondary stars are sufficiently cool, they likely display enhanced magnetic activity due to 
deep outer convective layers and rapid rotation. This magnetic mechanism may contribute to the period and light variations 
for systems with spectra later than F-type (Hall 1989).

CL Aur (GSC 2393-1455, HV 6886, TYC 2393-1455-1) was discovered to be a variable star by Hoffleit (1935) based on 
photographic plate estimates. Kurochkin (1951) presented the first (partial) photographic light curve of the star and 
the original light elements, Min. I = HJD 2,432,967.262 + 1.2443666$E$. The value of the period positions this object 
toward the short-period limit for Algols. The spectral type of the primary star was classified to be A0 
by G\"otz \& Wenzel (1968). Since then, times of minimum light have been published assiduously by numerous workers but, 
to our knowledge, a complete light curve and the fundamental parameters for the binary system have not been made so far. 
Changes of the orbital period have been considered by Heged\"us (1988) and Wolf et al. (1999). Heged\"us selected 
this system as a possible candidate for the study of apsidal motion. However, the later authors ruled out 
this possibility from CCD timings for primary and secondary eclipses. They suggested the cause of period variation to be 
a light-travel-time (LTT) effect due to the presence of a third body in the binary system.  Most recently, 
Wolf et al. (2007, hereafter W07) reported that a long-term period increase is superimposed on an LTT orbit with 
a period of $P_3 $=21.7 yrs, a semi-amplitude of $K$=0.014 d, and an eccentricity of $e$=0.32. 

In the Simbad data base\footnote {http://simbad.u-strasbg.fr/simbad/}, the system is described as an eclipsing binary of 
$\beta$ Lyr type. $BVJHK$ magnitues are listed for the star but these are from heterogeneous sources and are not mutually 
consistent. Part of this inconsistency arises because the magnitudes refer to different phases in the Keplerian cycle.
With the well-known transformations (ESA 1997), standard photometric values for CL Aur in the Johnson system were 
calculated to be $V=+$11.62 and $(B-V)=+$0.33 from the Tycho results. These refer to some unknown Keplerian phase.  
These are not consistent with those in Simbad presumably because of the large phase-locked variations in magnitude and 
color index of the binary. 

At present, CL Aur is known only as a neglected eclipsing system composed of an A-type primary and a cooler companion. 
In order to derive photometric solutions and to examine whether the W07 suggestion is appropriate for 
the orbital period change, we decided to obtain light curves with multiband photometry. In this paper, we present 
the first mutual analyses of the $O$-$C$ diagram and the light curves.

\section{CCD PHOTOMETRIC OBSERVATIONS}

New photometric observations of CL Aur were obtained using a SITe 2K CCD camera and a $BVRI$ filter set attached 
to the 61-cm reflector at Sobaeksan Optical Astronomy Observatory (SOAO) in Korea. The observations of 
the first season were made on 14 nights from 2003 November to 2004 March and those of the second season on 8 nights 
from 2004 December through 2005 February. The exposure times were 75$-$140 s for $B$, 45$-$85 s for $V$, 33$-$65 s 
for $R$, and 30$-$60 s for $I$, respectively, depending on weather conditions. The instrument and reduction method 
have been described by Lee et al. (2007) and a 2$\times$2 binning mode was selected. The nearby stars GSC 2393-1424 
and GSC 2393-1418, imaged on the chip at the same time as the variable, were selected as comparison and check stars, 
respectively. Coordinates and Tycho magnitudes for the three stars of interest are given in Table 1. 
The differential atmospheric extinction among the three stars is negligible within observational error. Measurements 
of the check and comparison stars indicate that the latter remained constant throughout the observing interval. 
The 1$\sigma$-values of the dispersions of the magnitude differences between them are about $\pm$0.01 mag 
for all bandpasses.  

A total of 2747 individual observations was obtained among the four bandpasses (711 in $B$, 693 in $V$, 685 in $R$, 
and 658 in $I$) and a sample of them is listed in Table 2. The light curves of CL Aur defined by our CCD photometry 
are plotted in Figure 1 as differential magnitudes {\it versus} orbital phase, which was computed according to 
the ephemeris for our hot-spot model determined later in this article with the Wilson-Devinney synthesis code 
(Wilson \& Devinney 1971, hereafter W-D).  The filled and open circles are the individual measures of the first and 
second observing seasons, respectively. Our light curves from the first observing season had not defined 
secondary eclipse adequately. Except formally for the $B$ bandpass, mean brightness differences between 
the two seasons (in the sense of Season 1 {\it minus} Season 2) are constant and smaller than the observational error 
of $\pm$0.01 mag: +0.013$\pm$0.014 mag for $B$, +0.007$\pm$0.016 mag for $V$, +0.001$\pm$0.017 mag for $R$, and 
$-$0.009$\pm$0.015 mag for $I$, respectively.  Although we show no figure illustrating variability of the color indices, 
their phase-locked variations are large and in the expected senses. These variations convey the idea of eclipses that 
are complete or very nearly so.

In addition, we observed several eclipse timings using a SBIG ST-8 CCD camera attached to the 35-cm reflector at 
the campus station of the Chungbuk National University Observatory (CbNUO) in Korea. The observations were made 
without a filter and reduced with the conventional IRAF package. The details of the CbNUO observations have been 
given by Kim et al. (2006).

\section{ORBITAL PERIOD STUDY}

We determined fifteen times of minimum light from all our CCD observations using the method of Kwee \& van Woerden 
(1956, hereafter KvW). These timings are listed in Table 3 together with all other CCD timings; the SOAO data
are weighted means from the observations in the $BVRI$ bandpasses. The second column gives the standard deviation of 
each timing. For the period study of CL Aur, 198 archival timings (16 plate, 101 visual, 8 photographic and 73 CCD) 
have been collected from the literature and from our measurements.  Most of the earlier timings were actually extracted 
from the data base published by Kreiner et al. (2001).  Because almost all but the CCD timings were published without 
error information, the following standard deviations were assigned to timing residuals based on observational method: 
$\pm$0.022 d for sky-patrol plate or photographic minima, and $\pm$0.007 d for visual minima. Relative weights were 
then calculated as the inverse squares of these values consistent with the errors and weights for the CCD timings.

In order to examine whether the period change of CL Aur can be produced by a quadratic {\it plus} LTT ephemeris 
as suggested by W07, we fitted all times of minimum light to that ephemeris form:
\begin{eqnarray}
C = T_0 + PE + AE^2 + \tau_3,
\end{eqnarray}
where $\tau_{3}$ symbolizes the LTT effect due to a third body (Irwin 1952, 1959) and includes five parameters 
($a_{12}\sin i_3$, $e$, $\omega$, $n$ and $T$). Here, $n$ and $T$ denote the Keplerian mean motion of the mass center 
of the eclipsing pair and the epoch of its periastron passage, respectively. In this analysis, 
the Levenberg-Marquart technique (Press et al. 1992) was applied to solve for the eight unknown parameters of 
the ephemeris. By using the third-body parameters of W07 as initial values, we obtained improved 
(in the sense of errors smaller than formerly) parameters for them and the results are listed in Table 4 together with 
other related quantities. In this table, $P_3$ and $K$ indicate the cycle length and semi-amplitude of the LTT orbit, 
respectively, and $f(M_{3})$ is the mass function of the system. Within errors, our LTT parameters are not significantly 
different from those of W07. The sample masses ($M_3 \sin i_3$) of the assumed third body in the table were calculated 
by using the absolute dimensions of the eclipsing pair presented in a later section. 

The $O$--$C$ diagram of CL Aur constructed with the linear terms in Table 4 is drawn in the top panel of Figure 2. 
The timings are marked by different symbols according to observational method and type of eclipse. The continuous curve 
and the dashed, parabolic one represent the full contribution and the quadratic term of the equation, respectively. 
The middle panel displays the residuals from the linear and quadratic terms, and the bottom panel the CCD residuals 
from the full ephemeris. These appear as $O$--$C_{\rm full}$ in the fourth column of Table 3.  As indicated by the figure, 
the quadratic {\it plus} LTT ephemeris gives a satisfactory representation of the ensemble of the residuals. 
In addition, another modulaton with a period of about five yrs and a semi-amplitude of about 0.0004 d {\it may} exist 
in the $O$--$C_{\rm full}$ residuals. A large number of future accurate timings is required before this can be tested 
at an acceptable level. If the orbit of the outer component is coplanar with that of the close eclipsing pair 
($i_3$=88$^\circ$.2), its mass is about $M_3$=0.83 M$_\odot$ and corresponds to a spectral type of K1--K2 for 
a normal main-sequence star. This would contribute about 0.9\% to the total light of the triple system, so it will be 
difficult to detect such a third light source from light-curve analysis. 

The quadratic term (A) of the ephemeris indicates a continuous period increase at a rate of 
+(1.4$\pm$0.2)$\times$10$^{-7}$ d yr$^{-1}$. Because CL Aur is a semi-detached system with the less massive and 
cool secondary component filling its inner Roche lobe (cf. Section 4), its Roche-geometry configuration permits 
mass transfer from the secondary to the more massive primary star, and it is conventional to ascribe such an increase 
to conservative mass transfer between the stars in the system. The calculated mass transfer rate is of the order of 
1.3$\times$10$^{-7}$ M$_\odot$ yr$^{-1}$, among the largest rates for Algol-type systems.

Alternatively, the 21.6 yr oscillation in the $O$--$C$ residuals could be caused by magnetic modulation due to 
an activity cycle in the convective envelope of the late-type star  (Applegate 1992, Lanza et al. 1998). 
The hot primary component of CL Aur likely has a radiative envelope as surmised from its spectral type while 
the less massive and cool secondary should have a shallow convective shell and at most weak magnetic activity. 
We applied the period ($P_3$) and amplitude ($K$) to Applegate's formulae and obtained model parameters for 
possible magnetic activity. The results are listed in Table 5, where the bolometric magnitude difference 
($\Delta m_{\rm rms}$) relative to the mean light level was obtained with equation (4) in the paper of 
Kim et al. (1997). Most of the tabulated values are close to those derived for several other binaries that appear to 
support Applegate's theory, but the light variation predicted from an active CL Aur secondary is at the upper limit 
of the theoretical value ($\Delta L/L_2 \sim 0.1$) proposed by him. Since our observations represent 
the first complete light curve, we cannot test whether such a variation has occurred in the past.  However, 
because the secondary star is expected not to be strongly magnetically active, we think the most probable explanation 
of the periodic oscillation to be the LTT effect due to a low-luminosity K-type tertiary companion.

\section{LIGHT-CURVE SYNTHESIS}

As shown in Figure 1, our observations clearly indicate that the light curve morphology of CL Aur is not $\beta$ Lyr type but 
rather very similar to that of Algol, its class prototype. To understand the geometrical structure and the physical parameters 
of the system, our $BVRI$ light curves were solved simultaneously in a manner similar to those for XX Cep (Lee et al. 2007) 
and GW Gem (Lee et al. 2009a) by using the latest version\footnote {ftp://ftp.astro.ufl.edu/pub/wilson/} of the W-D code 
and an extensive $q$-search procedure. The surface temperature of the primary star was fixed at $T_{1}$=9,420 K, according 
to its spectral type A0 and Harmanec's (1988) table. 
We had attempted to improve the spectral classification by obtaining high-resolution spectra with the echelle spectrograph 
attached to the 1.8-m telescope at Bohyunsan Optical Astronomical Observatory (BOAO) in Korea (Kim et al. 2002). 
These images were not well enough exposed to improve upon the literature value but indicated a similar result.
Initial bolometric and monochromatic limb-darkening coefficients were taken from the tables of van Hamme (1993) and were 
used together with the model atmosphere option. The $q$-search for modes 2, 3, 4 and 5 of the synthesizing code 
(Wilson \& Biermann 1976) converged and showed acceptable photometric solutions only for mode 5 
(semi-detached systems for which the secondary stars fill the inner Roche lobes). As displayed in Figure 3, 
the optimal solution is around $q$=0.60. This undertanding conforms to the sense of the mass transfer from 
the secondary component to the hotter, more massive primary star suggested by the period study. 

The $q$ value was treated as an adjustable parameter in all subsequent syntheses deriving binary parameters. The best result 
for unspotted photospheres is listed in columns (2) and (3) of Table 6 and the residuals from the analysis are plotted 
in the left panels of Figure 4, where phase-locked,  unmodelled light variations are indicated. Such features can be reasonably 
attributed to a hot spot on the surface of the primary as a result of the impact of the gas stream from the cooler, 
less massive secondary star. This situation is known for the Algol-type, semi-detached binary RZ Cas (Rodriguez et al. 2004). 

This interpretation does not exclude the possible existence of magnetic cool spots located on the surface of the late-type star. 
We therefore tested two spot models: a hot spot on the more massive primary star due to mass transfer and a cool spot 
on the secondary star caused by magnetic activity. In a formal sense, as shown by the entries on the last line of Table 6,
the hot spot model does improve the light-curve fit. Separate trials for a cool spot on the secondary star were not 
so successful as for the hot-spot model. Final results for all the light curves are given in columns (4) and (5) of 
Table 6 and the residuals from our hot-spot model are shown in the right panels of Figure 4. As seen in the figure, 
there is a slight improvement in the residuals for the spot model at about phase 0.87 in all panels compared to 
the unspotted model. This orbital phase is almost exactly where we would expect the effects of a gas stream to be evident. 
The hot spot agrees well with the concept of mass transfer from the secondary to the primary component inferred from 
the period analysis and from the relatively large size of the hot star.  In addition, there are small systematic differences 
between the two seasons for the light residuals, which means that mass transfer activity may be variable, as is sometimes reported  
for light curves of Algol binaries. In all these trials, a possible third light source ($\ell_{3}$) was considered but the results 
remained indistinguishable from 0.00 within their errors. We fixed $\ell_{3}$ to be 0.0 during the final light-curve analysis.

Because of the almost complete eclipses, the light curve determinacy for CL Aur is quite high. In idealized Roche geometry, 
the impact of the gas stream should be on the primary star's equator but this is not the case for CL Aur. We return to 
this detail later. It could also be true that both hot and cool spots exist and that our data are sufficient 
to isolate only the dominant one.  

A recent study by Lee et al. (2009b) indicates that the minimum epochs of the cool contact binary AR Boo have been 
systematically shifted by light asymmetries due to spot activity, as predicted by Maceroni \& van't Veer (1994). 
To check this possibility for this Algol-type system, we calculated the timings for each of our CL Aur eclipses 
with the W-D code. The results are listed in the second column of Table 7, together with the minimum times obtained 
by the KvW method for comparison. We can see that the differences among them  in column (4) of Table 7 are 
within the precision of each KvW minimum and the hot spot does not inflect those timings.  This agreement is 
doubtless due to the small temperature contrast of the spot against the photosphere and the small angular deviation
of the spot from the systemic line of centers.

A new class of $\delta$ Sct stars designated as oEA (oscillating EA) objects (Mkrtichian et al. 2004) has been identified 
as the (B)A-F spectral type mass-gaining components of Algol-type semi-detached systems. About half of them have been 
discovered through a photometric survey project run by a Korean group (cf. Kim et al. 2003) in order to search for 
A-type pulsating components in classical Algols (Soydugan et al. 2006). The oEA stars have pulsational amplitudes and 
periods similar to classical $\delta$ Sct stars but a different evolutionary scenario due to mass accretion. 
From its spectral type, the primary star of CL Aur would be a candidate for such pulsations. Therefore, we applied 
the discrete fourier transform program PERIOD04 (Lenz \& Breger 2005) to the light curve residuals from our binary models, 
but there was found no periodicity with a semi-amplitude larger than 3 mmag.

\section{ESTIMATED ABSOLUTE DIMENSIONS}

The foregoing presents a consistent picture of CL Aur in the sense that a determinate representation of the light curve 
has been achieved and that the secular period variability has a highly probable cause consistent with the light curve 
interpretation. Now it is necessary to look at further consequences of this description.

Absolute dimensions of CL Aur can be estimated from our photometric solutions with the hot-spot model in Table 6 and 
Harmanec's relation between spectral type and mass. By assuming the  primary star to be a normal main-sequence one 
with a spectral type of A0 V, the astrophysical parameters for the components were obtained to be those listed 
in Table 8, where the radii are the mean volume radii calculated from the tables of Mochnacki (1984). 
The luminosities and the bolometric magnitudes were computed by adopting $T_{\rm eff}$$_\odot$=5,780 K and 
$M_{\rm bol}$$_\odot$=+4.69. The bolometric corrections were obtained from the relation between $\log T$ and BC given 
by Kang et al. (2007). The intrinsic color of $(B-V)\rm_{0}=+$0.06 for the binary system was estimated from 
their calibration $(B-V)\rm_{1}$ {\it versus} $T_{1}$ of the primary star and from our computed light ratio; it leads 
to $E(B-V)$=$+$0.28. For lack of better evidence, it is prudent to assume that the mean system color index and 
interstellar reddening can be in error by as much as $\pm$0.05.  With this qualification and with the values of 
$V$ and $M_{\rm V,total}$ and the interstellar extinction of $A_{\rm V}$=3.1$E(B-V)$, we have calculated 
a nominal distance to the system of about 1,150 pc and an accidental error associated this distance of about 5 \%.  
This result is also misleading for another reason: one does not know the phase at which the Tycho measures were made 
so there is no assurance that they refer to maximum light. The total error on the distance determination may well be 20\%.
 
From the estimated parameters of CL Aur, it is possible to consider the evolutionary state of the eclipsing system 
in mass-radius and mass-luminosity diagrams. The primary star lies in the main-sequence band between the ZAMS 
and the TAMS, while the secondary is larger and brighter than expected for its mass. The locations of the components 
in the Hertzsprung-Russell (HR) diagram are shown in Figure 5, together with single-star evolutionary tracks having 
masses of 2.20 $M_\odot$ and 1.35 $M_\odot$ and solar metallicity (Girardi et al. 2000). There also appear 
in the figure theoretical mass-conserving evolutionary tracks for model donor (3.0 $M_\odot$, the present secondary) 
and gainer (1.2 $M_\odot$) stars belonging to a system with a total mass of 4.2 $M_\odot$ and an initial orbital period 
of 1.0 d calculated by the Brussels group\footnote {http://we.vub.ac.be/astrofys/} (De Loore \& van Rensbergen, 2005).
Although these tracks are not for a system with the same mass ratio and total mass as CL Aur (3.6 $M_\odot$), 
the two components can be scaled to fit the concept of a similar conservative model quite well, indicating that 
the binary system may still be undergoing mass transfer as a result of case A evolution. The present age of the system 
is estimated to be about 0.24 Gyr.

\section{DISCUSSION AND CONCLUSIONS}

We next probed the credibility of the hot (inferentially an impact) spot model motivated largely by two recognitions:  
(a) the two arrays of light curve residuals in Figure 4 do not appear very different one from another although formally 
the light curve fitting is improved by the assumption of a spot and (b) the center of the spot is sensibly distant 
from the stellar equator and orbital plane.  

The first matter is the consequence of the envelope of the residuals responding weakly to the spot modeling although 
the residuals of small absolute value did respond well with considerably reduced values on average.  This can only be 
construed as a situation in which there exist residuals of noise much greater than most of the rest of them. 
In principle, these residuals could be produced by a combination of physical causes such as magnetic activity from 
the cool secondary and a change in the mass transfer rate. However, the actual noise level is not significantly larger 
than that ($\pm$0.01 mag) of SOAO data made during the last few observing seasons and probably can be traced to 
marginal observing conditions on some nights. 

The location of the spot has more physical interest.  First, an extensive set of restricted 3-body calculations examined 
the longitude coordinate of the spot on the hot primary for a range (0.0000003 km s$^{-1}$ -- 450 km s$^{-1}$) of 
initial velocities streaming from the L$_1$ point as shown in the first column of Table 9. This range obviously includes 
the thermal velocity in the cool star envelope with the lower limit of the range essentially that of free-fall. 
For the greatest extent of this velocity range the impact spot is close (within 21$^\circ$) to the line of centers.  
In the hydrodynamic computations of Lubow \& Shu (1975), just such an effect appears with the streaming gas deflected 
about 20$^\circ$ from the line of centers, not nearly enough to avoid contact with the large, detached primary. 
From this point of view, the credibility of the concept of impact is therefore high, and conservative mass transfer 
of the streaming material must be considered likely.  These calculations also make it difficult to imagine 
any stable accretion structures around the hot star but they invariably led to an orbital-plane spot rather than 
the modeled one of co-latitude very different from 90$^\circ$. Another suite of calculations, also given in Table 9, 
showed the expected result: inital non-zero-velocity $z$-components in the streaming gas led to impacts away from 
the hot stellar equator and these did not have to be large in order to fall 15--20$^\circ$ away from the equator. 
Is this result to be taken as evidence that the spot really exists at its modeled location or is it just 
a dynamical truism and doesn't verify the spot existence at all?  The resolution of this quandary requires 
a mechanism to move the gas appropriately and three possible ones come to mind.  (a) Turbulence in the gas moving 
from the L$_1$-point caused a concentration of it to fall at the modeled position during the two observing seasons. 
(b) A weak magnetic field seated in the cool star had at least one component channeling the ionized fraction of 
the streaming gas to the modeled impact spot. (c) The solution for the spot is possibly not so unique as W-D indicates.  
For want of independent evidence, each of these is 
an unprovable hypothesis.

More information does exist, however.  We calculated the impact luminosity from the stream at the rate of mass transfer 
given by the interpretation of the $O$--$C$ diagram.  For the same range (and in the same sense as above) of stream velocity, 
the impact luminosity varied between 3.2 $L_\odot$ and 0.7 $L_\odot$. If the impact energy is partitioned between 
virialization into the gainer star and spot luminosity as Hilditch (1989) proposes, the least luminous output from 
a spot should therefore be about 0.35 $L_\odot$. This is to be compared to the spot luminosity from its black body 
W-D model of about 4$\times10^{-5}$ $L_\odot$.  The magnitude of the discrepancy means either (a) that conversion 
from kinetic to luminous energy is very inefficient or (b) that the mass-motions rate indicated by the $O$--$C$ diagram 
is not restricted to material leaving the L$_1$ point or (c) some combination of these two ideas or (d) that the concept 
of a spot is itself incorrect.  For (a) to be a realistic interpretation, deep penetration of the impacting gas into 
the hot star is required so that most of the impacting gas becomes thermalized in the hot star. It is also necessary 
to postulate an envelope circulation pattern that re-surfaces at the modeled spot position for the residual gas that 
has not been thermalized.  If (b) is to be entertained seriously, there must be substantial mass lost by evaporation 
from the entire Roche lobe that is the photosphere of the cool star.  This demands only a small ($\sim 10^{-4}$) 
contribution to the $O$--$C$ diagram from the gas leaving L$_1$ with the majority of the mass and angular momentum loss 
contributed by that departing the rest of the cool star's surface. On a small scale the transfer would be 
conservative as usual but large scale mass loss would be the dominant cause of the $A$-term in the ephemeris. 
Hypothesis (d) would be the most economical interpretation but conflicts with the impersonal syntheses of 
the light curves. At this time, we favor a combination of interpretations (a) and (b).

In summary, our study of the orbital period and the light curves reveals that CL Aur is a classical Algol-type interacting system 
with the less massive and cool secondary star filling its inner Roche lobe. The possibility of a hot-spot model due to 
impact of streaming gas onto the hot star has led to confusing difficulties that we have resolved only tentatively.  
High-resolution, near-IR spectroscopy should reveal the lines of the secondary star and lead to accurate absolute parameters 
to replace our estimates. Moreover, there is also the possibility of obtaining direct evidence of mass-transfer activity, 
such as complex and variable line profiles of H$\alpha$ or variations in the strength of the O I 7774 absorption line.
The evolutionary status of the system will then be more convincingly in hand.  

The more general reality of things is that there are many short-period Algols which have not been studied so deeply 
as this present work has examined CL Aur. It would make a significant advance if more of these - 
having different algebraic signs for the secular term of the period variability - could be brought to the same level 
of knowledge as the present binary. For instance, should all such binaries with $A >$ 0 require a spot on the hot star 
near the line of centers, there would be major support for the reasoning concerning mass movements.  Should it also happen 
that the spot luminosity was consistently found to be lower than the kinetic impact required for 50\% energy conversion, 
there would be good reason to believe that the majority of the mass lost from the cool secondaries is, in fact, 
lost to the systems and not transferred conservatively. The short-period binaries found to require mode 4 
(semi-detached systems with the primary stars filling its inner Roche lobes) for their representations would be 
expected, then, to show small-scale mass transfer to the secondaries and systemic mass loss from the  primaries.  
Much valuable observational and modeling work remains.

\acknowledgments{ }
We would like to thank the staff of the Chungbuk National University Observatory for assistance with our observations, 
and Mr. Jae-Rim Koo for the spectroscopic observations of CL Aur.  We appreciate the careful reading and valuable comments 
of the anonymous referee.We have used the Simbad data base maintained at CDS many times and appreciate its availability.  
This work has been done as part of a cooperative project between Chungbuk National University and the Korea Astronomy 
and Space Science Institute.  C.-H. K. was supported by the Korea Research Foundation (KRF) Grant funded by 
the Korea government (MEST) (No. 2009-0069330).

\clearpage
\begin{figure}
 \includegraphics[]{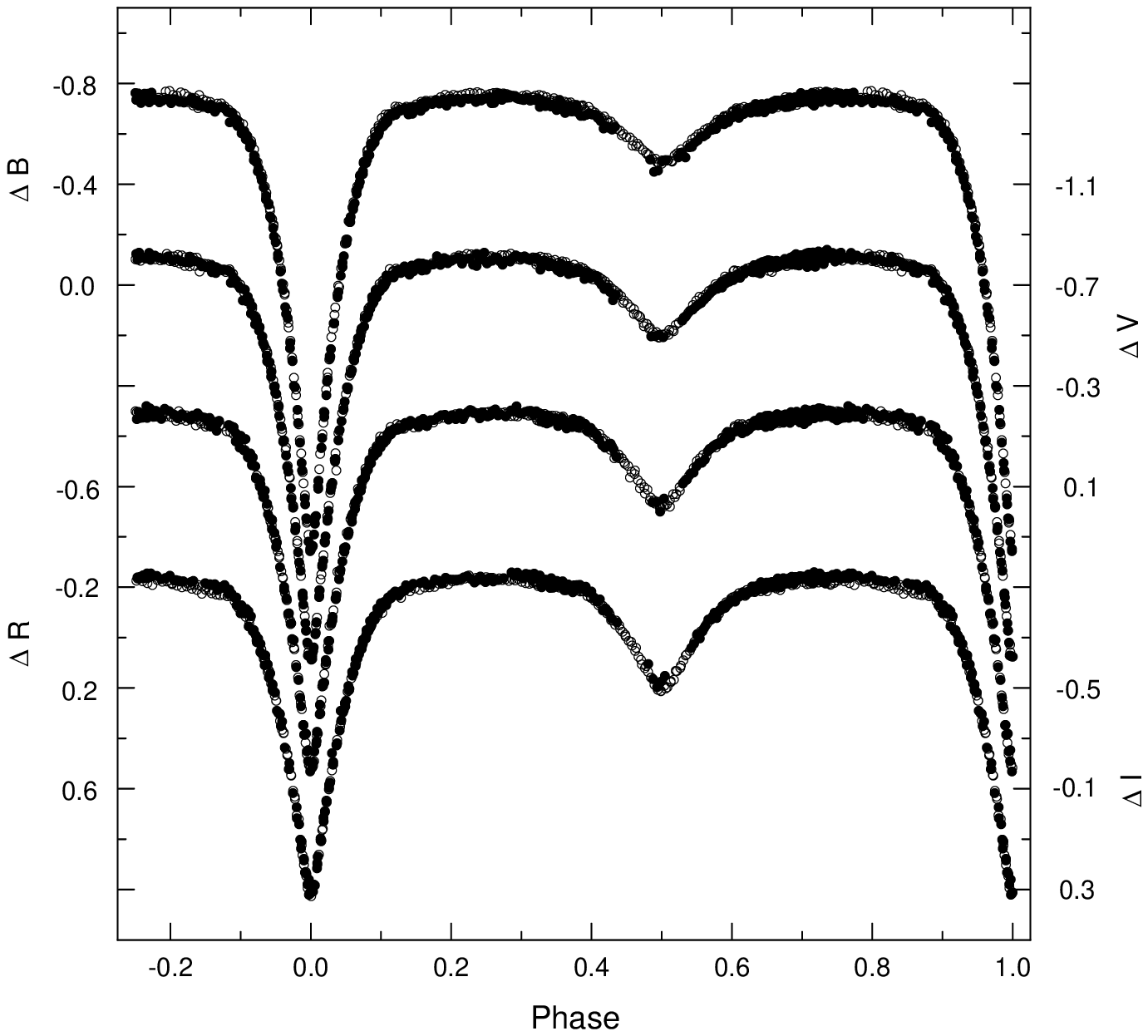}
 \caption{SOAO light curves of CL Aur in the $B$, $V$, $R$, and $I$ bandpasses. The filled and open circles are 
 the individual measures of the first (2003 Nov$-$2004 Mar) and second (2004 Dec$-$2005 Feb) seasons, respectively. 
 Because of the high density of the points, many of the 2004 measures cannot be seen individually.}
 \label{Fig1}
\end{figure}

\begin{figure}
 \includegraphics[]{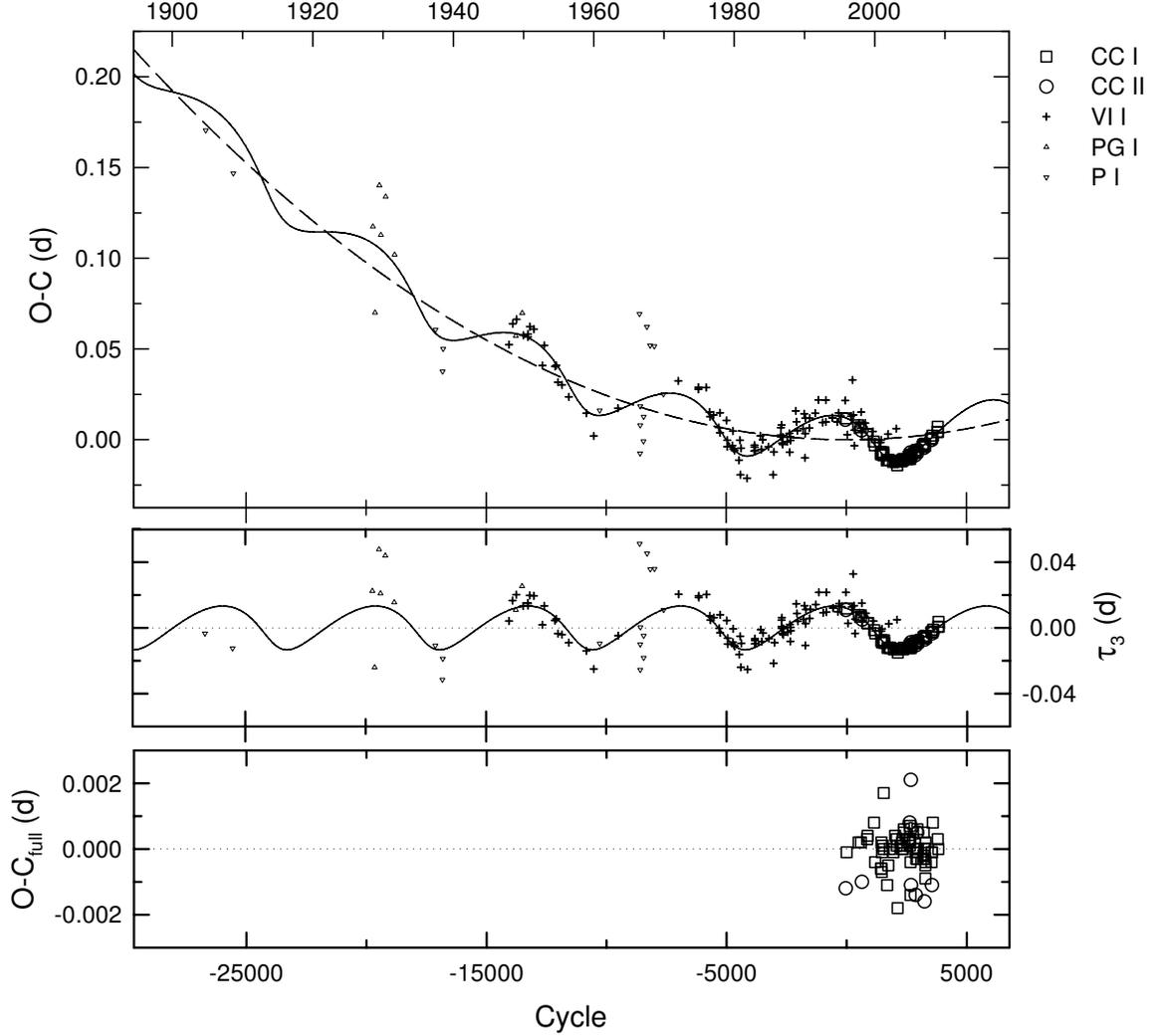}
 \caption{The $O$--$C$ diagram of CL Aur constructed with the linear terms in Table 4.  In the top panel, the continuous curve 
 and the dashed, parabolic one represent the full contribution and the quadratic term of the equation, respectively. 
 The middle panel displays the residuals from the linear and quadratic terms, and the bottom panel the CCD residuals from 
 the complete ephemeris. CC, VI, PG, and P denote CCD, visual, photographic, and photographic plate minima, respectively.}
\label{Fig2}
\end{figure}

\begin{figure}
 \includegraphics[]{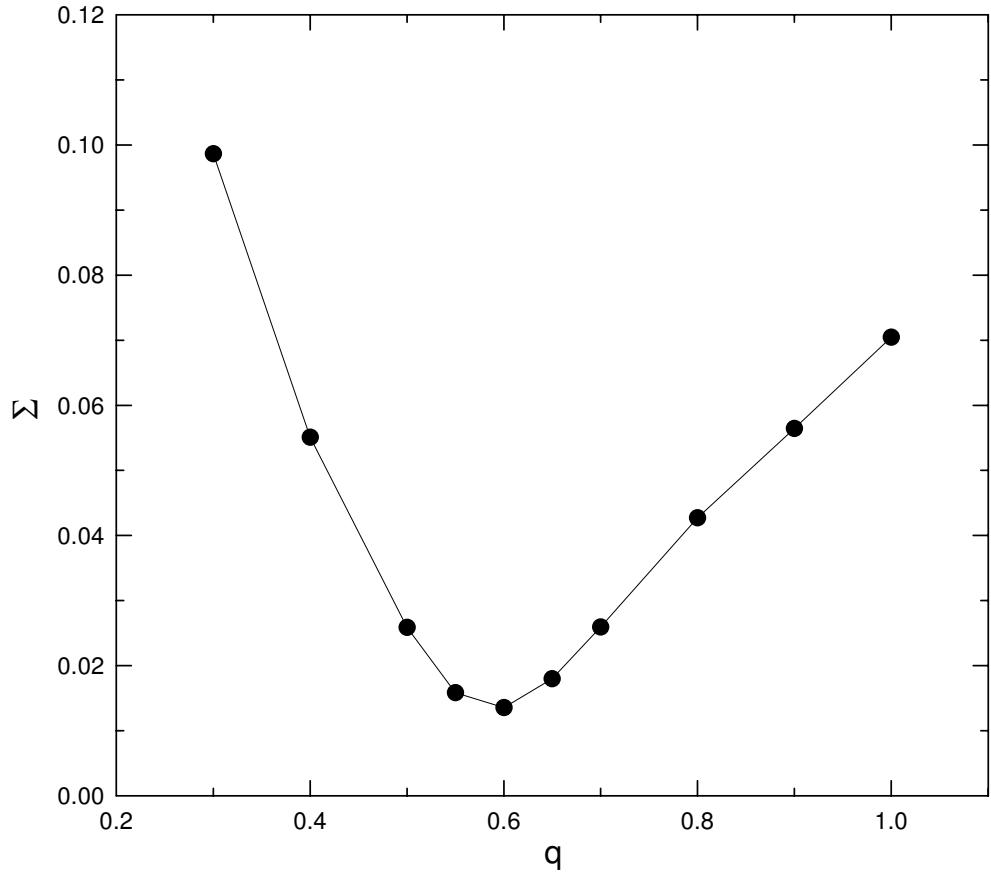}
 \caption{The behavior of $\Sigma$ (the sum of the residuals squared) of CL Aur as a function of mass ratio $q$, showing 
 a minimum value near $q$=0.60 for Mode 5 of the W-D code.}
\label{Fig3}
\end{figure}

\begin{figure}
 \includegraphics[]{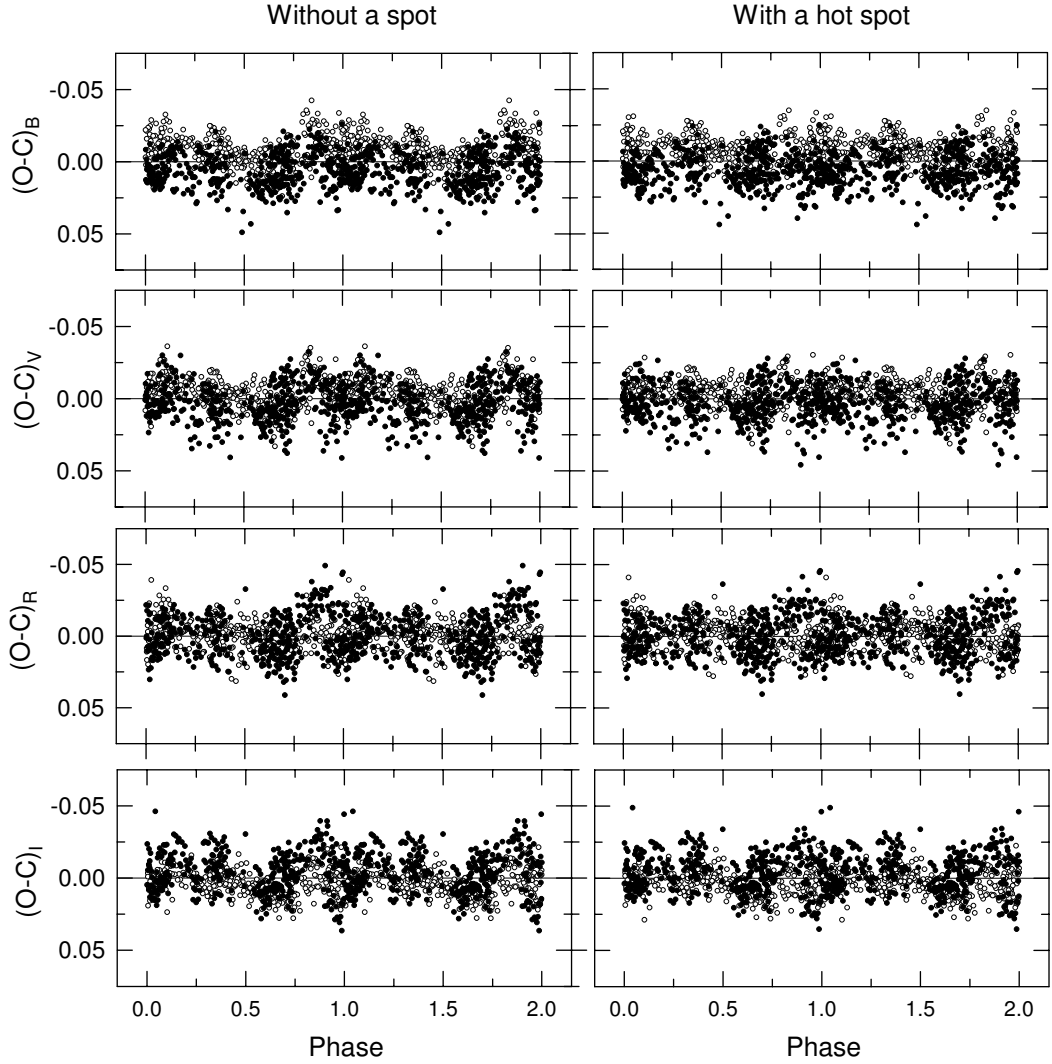}
 \caption{The magnitude residuals in $B$, $V$, $R$ and $I$ corresponding to two binary models in columns (2)-(5) of 
 Table 6: without (left panels) and with (right panels) a hot spot on the primary component. The symbols are plotted 
 in the same sense as Figure 1. }
\label{Fig4}
\end{figure}

\begin{figure}
 \includegraphics[]{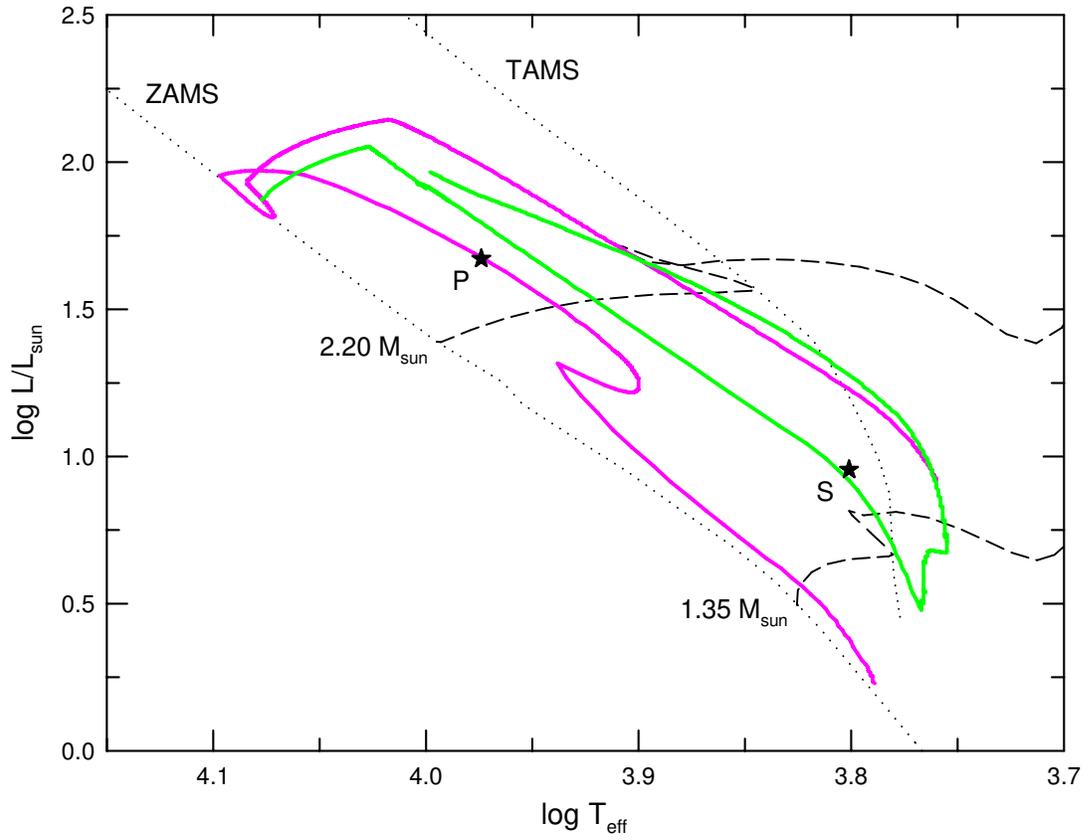}
 \caption{The HR diagram of the primary (P) and secondary (S) stars of CL Aur. The dashed lines correspond to 
 single star evolutionary tracks for the present masses of both components. The green- and pink-colored lines 
 represent theoretical tracks for closely similar donor and gainer stars, respectively. See the text for details. }
\label{Fig5}
\end{figure}

\clearpage
\begin{deluxetable}{cccccc}
\tablewidth{0pt} 
\tablecaption{Coordinates and photometric data for the program stars.}
\tablehead{
\colhead{Star} & \colhead{GSC} & \colhead{RA (J2000)} & \colhead{DEC (J2000)} & $V\rm_T^\dagger$  &  $(B-V)\rm_T^\dagger$}             
\startdata                                                                                                                             
CL Aurigae     &  2393-1455     &  05$^{\rm h}$12$^{\rm m}$54$\fs19$  &  +33$^{\circ}$30${\rm '}$28$\farcs$4  & $+$11.65  &  $+$0.39   \\
Comparison     &  2393-1424     &  05$^{\rm h}$13$^{\rm m}$27$\fs48$  &  +33$^{\circ}$26${\rm '}$46$\farcs$3  &  --     &  --        \\
Check          &  2393-1418     &  05$^{\rm h}$12$^{\rm m}$19$\fs08$  &  +33$^{\circ}$26${\rm '}$31$\farcs$5  & $+$12.30  &  $+$0.14     
\enddata
\tablenotetext{\dagger}{From the Tycho-2 Catalogue (H\o g et al. 2000).}
\end{deluxetable}

\begin{deluxetable}{crcrcrcr}
\tabletypesize{\small}
\tablewidth{0pt} 
\tablecaption{SOAO CCD photometric observations of CL Aur.}
\tablehead{
\colhead{HJD} & \colhead{$\Delta B$} & \colhead{HJD} & \colhead{$\Delta V$} & \colhead{HJD} & \colhead{$\Delta R$} & \colhead{HJD} & \colhead{$\Delta I$}
}
\startdata
2,452,974.23806 & 0.707  &  2,452,974.23966 & 0.525  &  2,452,974.24094 &  0.348  &  2,452,974.23583 &  0.104   \\
2,452,974.24426 & 0.872  &  2,452,974.24587 & 0.673  &  2,452,974.24716 &  0.452  &  2,452,974.24215 &  0.179   \\
2,452,974.24997 & 1.020  &  2,452,974.25159 & 0.775  &  2,452,974.25288 &  0.532  &  2,452,974.24832 &  0.280   \\
2,452,974.25568 & 1.042  &  2,452,974.25732 & 0.760  &  2,452,974.25861 &  0.496  &  2,452,974.25404 &  0.312   \\
2,452,974.26142 & 0.949  &  2,452,974.26304 & 0.682  &  2,452,974.26433 &  0.412  &  2,452,974.25977 &  0.281   \\
2,452,974.26721 & 0.800  &  2,452,974.26882 & 0.540  &  2,452,974.27010 &  0.305  &  2,452,974.26550 &  0.196   \\
2,452,974.27288 & 0.646  &  2,452,974.27450 & 0.420  &  2,452,974.27579 &  0.193  &  2,452,974.27127 &  0.106   \\
2,452,974.27860 & 0.507  &  2,452,974.28020 & 0.299  &  2,452,974.28149 &  0.098  &  2,452,974.27697 &  0.030   \\
2,452,974.28423 & 0.377  &  2,452,974.28585 & 0.198  &  2,452,974.28714 &  0.014  &  2,452,974.28266 & -0.049   \\
2,452,974.29000 & 0.255  &  2,452,974.29161 & 0.088  &  2,452,974.29289 & -0.082  &  2,452,974.28831 & -0.123   \\
\enddata
\tablecomments{This table is available in its entirety in machine-readable and Virtual Observatory (VO) forms 
in the online journal and also at the Web page (http://binary.cbnu.ac.kr/bbs/zboard.php?id=lab\_photometry). 
A portion is shown here for guidance regarding its form and content.}
\end{deluxetable}

\begin{deluxetable}{llrrcl}
\tablewidth{0pt} 
\tablecaption{Observed CCD times of minimum light for CL Aur.}
\tablehead{
\colhead{HJD} & Error & \colhead{Epoch} & \colhead{$O$--$C_{\rm full}$} & \colhead{Min} & \colhead{References}
}
\startdata
2,450,044.3958 &  $\pm$0.0007  &   -42.5  &  -0.00122  &  II  &  Wolf et al. (1999)            \\
2,450,097.2826 &  $\pm$0.0002  &     0.0  &  -0.00014  &  I   &  Wolf et al. (1999)            \\
2,450,714.4888 &  $\pm$0.0006  &   496.0  &   0.00020  &  I   &  Wolf et al. (1999)            \\
2,450,831.4589 &  $\pm$0.0001  &   590.0  &   0.00019  &  I   &  Wolf et al. (1999)            \\
2,450,884.3431 &  $\pm$0.0003  &   632.5  &  -0.00100  &  II  &  Wolf et al. (1999)            \\
2,451,157.4815 &  $\pm$0.0006  &   852.0  &   0.00027  &  I   &  Wolf et al. (1999)            \\
2,451,177.3914 &  $\pm$0.0001  &   868.0  &   0.00043  &  I   &  Wolf et al. (1999)            \\
2,451,508.391  &  $\pm$0.0030  &  1134.0  &   0.00080  &  I   &  Bl\"attler (2000)             \\
2,451,570.6077 &  $\pm$0.0003  &  1184.0  &  -0.00037  &  I   &  Baldwin \& Samolyk(2004)      \\
2,451,880.4526 &  $\pm$0.0002  &  1433.0  &  -0.00061  &  I   &  Br\'at et al. (2009)          \\
2,451,901.6067 &  $\pm$0.0002  &  1450.0  &  -0.00062  &  I   &  Wolf et al. (2007)            \\
2,451,901.6075 &  $\pm$0.0003  &  1450.0  &   0.00018  &  I   &  Br\'at et al. (2009)          \\
2,451,921.5164 &  $\pm$0.0002  &  1466.0  &  -0.00068  &  I   &  Agerer \& H\"ubscher (2002)   \\
2,451,925.2501 &  $\pm$0.0021  &  1469.0  &  -0.00006  &  I   &  Br\'at et al. (2007)          \\
2,451,956.3593 &  $\pm$0.0015  &  1494.0  &   0.00014  &  I   &  Br\'at et al. (2007)          \\
2,452,013.5998 &  $\pm$0.0002  &  1540.0  &   0.00004  &  I   &  Baldwin \& Samolyk(2004)      \\
2,452,017.3345 &  $\pm$0.0003  &  1543.0  &   0.00165  &  I   &  Wolf et al. (2007)            \\
2,452,196.5200 &  $\pm$0.0003  &  1687.0  &  -0.00113  &  I   &  Diethelm (2001)               \\
2,452,252.5171 &  $\pm$0.0001  &  1732.0  &  -0.00047  &  I   &  Wolf et al. (2007)            \\
2,452,333.4014 &  $\pm$0.0001  &  1797.0  &  -0.00002  &  I   &  Wolf et al. (2007)            \\
2,452,522.5455 &  $\pm$0.0001  &  1949.0  &  -0.00013  &  I   &  Wolf et al. (2007)            \\
2,452,609.6521 &  $\pm$0.0001  &  2019.0  &   0.00038  &  I   &  Baldwin \& Samolyk(2004)      \\
2,452,684.3143 &  $\pm$0.0003  &  2079.0  &   0.00012  &  I   &  Wolf et al. (2007)            \\
2,452,690.5364 &  $\pm$0.0002  &  2084.0  &   0.00034  &  I   &  Baldwin \& Samolyk(2004)      \\
2,452,731.5986 &  $\pm$0.0007  &  2117.0  &  -0.00185  &  I   &  Baldwin \& Samolyk(2004)      \\
2,452,899.5915 &  $\pm$0.0002  &  2252.0  &   0.00012  &  I   &  Wolf et al. (2007)            \\
2,452,964.2991 &  $\pm$0.0001  &  2304.0  &   0.00001  &  I   &  Wolf et al. (2007)            \\
2,452,974.2543 &  $\pm$0.0002  &  2312.0  &   0.00017  &  I   &  This paper (SOAO)             \\
2,452,986.6982 &  $\pm$0.0002  &  2322.0  &   0.00028  &  I   &  Baldwin \& Samolyk(2004)      \\
2,453,019.0518 &  $\pm$0.0001  &  2348.0  &   0.00000  &  I   &  This paper (SOAO)             \\
2,453,055.1393 &  $\pm$0.0002  &  2377.0  &   0.00046  &  I   &  This paper (SOAO)             \\
2,453,062.6057 &  $\pm$0.0002  &  2383.0  &   0.00058  &  I   &  Baldwin \& Samolyk(2004)      \\
2,453,323.9256 &  $\pm$0.0001  &  2593.0  &   0.00022  &  I   &  Baldwin \& Samolyk (2007)     \\
2,453,350.0577 &  $\pm$0.0001  &  2614.0  &   0.00027  &  I   &  This paper (SOAO)             \\
2,453,353.1692 &  $\pm$0.0006  &  2616.5  &   0.00081  &  II  &  This paper (SOAO)             \\
2,453,360.0128 &  $\pm$0.0001  &  2622.0  &   0.00030  &  I   &  This paper (SOAO)             \\
2,453,361.2570 &  $\pm$0.0002  &  2623.0  &   0.00011  &  I   &  This paper (SOAO)             \\
2,453,387.3896 &  $\pm$0.0005  &  2644.0  &   0.00065  &  I   &  H\"ubscher et al. (2005)      \\
2,453,388.6336 &  $\pm$0.0001  &  2645.0  &   0.00027  &  I   &  Smith \& Caton (2007)         \\
2,453,407.2977 &  $\pm$0.0006  &  2660.0  &  -0.00139  &  I   &  H\"ubscher et al. (2005)      \\
2,453,407.2987 &  $\pm$0.0035  &  2660.0  &  -0.00039  &  I   &  H\"ubscher et al. (2005)      \\
2,453,410.4121 &  $\pm$0.0003  &  2662.5  &   0.00205  &  II  &  H\"ubscher et al. (2005)      \\
2,453,414.1438 &  $\pm$0.0009  &  2665.5  &   0.00060  &  II  &  This paper (SOAO)             \\
2,453,425.3416 &  $\pm$0.0001  &  2674.5  &  -0.00106  &  II  &  Wolf et al. (2007)            \\
2,453,641.2436 &  $\pm$0.0003  &  2848.0  &   0.00016  &  I   &  This paper (CbNUO)            \\
2,453,675.4626 &  $\pm$0.0003  &  2875.5  &  -0.00144  &  II  &  B\'ir\'o et al. (2007)        \\
2,453,694.7519 &  $\pm$0.0002  &  2891.0  &  -0.00012  &  I   &  Baldwin \& Samolyk (2007)     \\
2,453,704.7068 &  $\pm$0.0001  &  2899.0  &  -0.00031  &  I   &  Nelson (2006)                 \\
2,453,713.4178 &  $\pm$0.0001  &  2906.0  &  -0.00001  &  I   &  Wolf et al. (2007)            \\
2,453,746.3945 &  $\pm$0.0002  &  2932.5  &   0.00047  &  II  &  Wolf et al. (2007)            \\
2,453,764.4382 &  $\pm$0.0005  &  2947.0  &   0.00057  &  I   &  H\"ubscher et al. (2006)      \\
2,453,769.4149 &  $\pm$0.0001  &  2951.0  &  -0.00027  &  I   &  Wolf et al. (2007)            \\
2,453,793.0585 &  $\pm$0.0001  &  2970.0  &  -0.00001  &  I   &  This paper (CbNUO)            \\
2,454,054.3794 &  $\pm$0.0001  &  3180.0  &  -0.00029  &  I   &  Do\u{g}ru et al. (2007)       \\
2,454,070.5565 &  $\pm$0.0002  &  3193.0  &  -0.00022  &  I   &  Wolf et al. (2007)            \\
2,454,084.2455 &  $\pm$0.0002  &  3204.0  &   0.00052  &  I   &  This paper (CbNUO)            \\
2,454,085.4892 &  $\pm$0.0006  &  3205.0  &  -0.00017  &  I   &  H\"ubscher \& Walter (2007)   \\
2,454,115.3547 &  $\pm$0.0010  &  3229.0  &   0.00004  &  I   &  H\"ubscher (2007)             \\
2,454,120.9528 &  $\pm$0.0005  &  3233.5  &  -0.00160  &  II  &  This paper (CbNUO)            \\
2,454,141.4868 &  $\pm$0.0002  &  3250.0  &   0.00002  &  I   &  Wolf et al. (2007)            \\
2,454,152.6860 &  $\pm$0.0001  &  3259.0  &  -0.00027  &  I   &  Baldwin \& Samolyk (2007)     \\
2,454,171.3512 &  $\pm$0.0009  &  3274.0  &  -0.00088  &  I   &  H\"ubscher (2007)             \\
2,454,171.3516 &  $\pm$0.0001  &  3274.0  &  -0.00048  &  I   &  Wolf et al. (2007)            \\
2,454,176.3298 &  $\pm$0.0001  &  3278.0  &   0.00018  &  I   &  Wolf et al. (2007)            \\
2,454,186.2843 &  $\pm$0.0002  &  3286.0  &  -0.00042  &  I   &  Wolf et al. (2007)            \\
2,454,487.4260 &  $\pm$0.0001  &  3528.0  &  -0.00045  &  I   &  Borkovits et al. (2008)       \\
2,454,501.11462&  $\pm$0.00007 &  3539.0  &  -0.00009  &  I   &  This paper (CbNUO)            \\
2,454,506.09212&  $\pm$0.00007 &  3543.0  &  -0.00014  &  I   &  This paper (CbNUO)            \\
2,454,510.4465 &  $\pm$0.0003  &  3546.5  &  -0.00112  &  II  &  Borkovits et al. (2008)       \\
2,454,558.3573 &  $\pm$0.0003  &  3585.0  &   0.00077  &  I   &  Br\'at et al. (2008)          \\
2,454,815.94500&  $\pm$0.00007 &  3792.0  &   0.00029  &  I   &  This paper (CbNUO)            \\
2,454,834.6136 &  $\pm$0.0001  &  3807.0  &   0.00308  &  I   &  Br\'at et al. (2009)          \\
2,454,843.3212 &  $\pm$0.0002  &  3814.0  &  -0.00003  &  I   &  Br\'at et al. (2009)          \\
\enddata
\end{deluxetable}

\begin{deluxetable}{lcc}
\tablewidth{0pt}
\tablecaption{Parameters for the quadratic {\it plus} LTT ephemeris of CL Aur.}
\tablehead{
\colhead{Parameter}               &  \colhead{Value}                   &  \colhead{Unit}          
}                                                                                                  
\startdata                                                                                         
$T_0$                             &  2,450,097,27082$\pm$0.00046        &  HJD                   \\
$P$                               &  1.24437498$\pm$0.00000017          &  d                     \\
$A$                               &  (2.44$\pm$0.34)$\times 10^{-10}$   &  d                     \\
$a_{12}\sin i_{3}$                &  2.38$\pm$0.12                      &  AU                    \\
$\omega$                          &  218.9$\pm$2.7                      &  deg                   \\
$e$                               &  0.337$\pm$0.053                    &                        \\
$n$                               &  0.04556$\pm$0.00029                &  deg d$^{-1}$          \\
$T$                               &  2,444,072$\pm$56                   &  HJD                   \\
$P_{3}$                           &  21.63$\pm$0.14                     &  yr                    \\
$K$                               &  0.01330$\pm$0.00069                &  d                     \\
$f(M_{3})$                        &  0.0290$\pm$0.0015                  &  $M_\odot$             \\
$M_3$ ($i_{3}$=90 deg)$^\dagger$  &  0.83                               &  $M_\odot$             \\
$M_3$ ($i_{3}$=60 deg)$^\dagger$  &  0.98                               &  $M_\odot$             \\
$M_3$ ($i_{3}$=30 deg)$^\dagger$  &  1.92                               &  $M_\odot$             \\
$dP$/$dt$                         &  (1.43$\pm$0.20)$\times 10^{-7}$    &  d yr$^{-1}$           \\
$dM$/$dt$                         &  1.30$\times 10^{-7}$               &  $M_\odot$ yr$^{-1}$   \\
\enddata
\tablenotetext{\dagger}{Masses of the third body for different values of $i_{3}$.}
\end{deluxetable}

\begin{deluxetable}{lcc}
\tablewidth{0pt}
\tablecaption{Applegate-model parameters for the cool secondary of CL Aur.}
\tablehead{
\colhead{Parameter}       & \colhead{Value}         & \colhead{Unit}
}
\startdata
$\Delta P$                &  1.1372                 &  s                     \\
$\Delta P/P$              &  $1.06\times10^{-5}$    &                        \\
$\Delta Q$                &  ${8.49\times10^{50}}$  &  g cm$^2$              \\
$\Delta J$                &  ${1.46\times10^{48}}$  &  g cm$^{2}$ s$^{-1}$   \\
$I_{\rm s}$               &  ${5.47\times10^{54}}$  &  g cm$^{2}$            \\
$\Delta \Omega$           &  ${2.68\times10^{-7}}$  &  s$^{-1}$              \\
$\Delta \Omega / \Omega$  &  ${4.59\times10^{-3}}$  &                        \\
$\Delta E$                &  ${7.85\times10^{41}}$  &  erg                   \\
$\Delta L_{\rm rms}$      &  ${3.61\times10^{33}}$  &  erg s$^{-1}$          \\
                          &  0.926                  &  $L_\odot$             \\
                          &  0.103                  &  $L_2$                 \\
$\Delta m_{\rm rms}$      &  $\pm$0.018             &  mag                   \\
$B$                       &  8,900                  &  G                     \\
\enddata
\end{deluxetable}

\begin{deluxetable}{lccccc}
\tablewidth{0pt} 
\tablecaption{Photometric solutions of CL Aur.}
\tablehead{
\colhead{Parameter}             & \multicolumn{2}{c}{Without spot}                 && \multicolumn{2}{c}{With spot}                    \\ [1.0mm] \cline{2-3} \cline{5-6} \\[-2.0ex]
                                & \colhead{Primary} & \colhead{Secondary}          && \colhead{Primary} & \colhead{Secondary}
}                                                                                  
\startdata                                                                         
$T_0$ (HJD)                     & \multicolumn{2}{c}{2,452,974.25441$\pm$0.00006}  && \multicolumn{2}{c}{2452974.25438$\pm$0.00005}    \\ 
$P$ (d)                         & \multicolumn{2}{c}{1.24438169$\pm$0.00000025}    && \multicolumn{2}{c}{1.24438175$\pm$0.00000025}    \\ 
$q$                             & \multicolumn{2}{c}{0.5944(55)}                   && \multicolumn{2}{c}{0.6023(49)}                   \\
$i$ (deg)                       & \multicolumn{2}{c}{88.15(11)}                    && \multicolumn{2}{c}{88.21(12)}                    \\
$T$ (K)                         & 9,420             & 6,315(46)                    && 9,420             & 6,323(44)                    \\
$\Omega$                        & 3.569(14)         & 3.053                        && 3.563(13)         & 3.068                        \\
$A$                             & 1.0               & 0.5                          && 1.0               & 0.5                          \\
$g$                             & 1.0               & 0.32                         && 1.0               & 0.32                         \\
$X$                             & 0.612             & 0.478                        && 0.612             & 0.478                        \\
$x_{B}$                         & 0.601(28)         & 0.375(103)                   && 0.585(27)         & 0.368(101)                   \\
$x_{V}$                         & 0.515(30)         & 0.539(70)                    && 0.511(27)         & 0.538(68)                    \\
$x_{R}$                         & 0.408(31)         & 0.528(54)                    && 0.409(28)         & 0.529(53)                    \\
$x_{I}$                         & 0.329(30)         & 0.397(45)                    && 0.332(28)         & 0.398(44)                    \\
$L/(L_1+L_2)_{B}$               & 0.8629(24)        & 0.1371                       && 0.8630(22)        & 0.1370                       \\
$L/(L_1+L_2)_{V}$               & 0.8163(24)        & 0.1837                       && 0.8163(22)        & 0.1837                       \\
$L/(L_1+L_2)_{R}$               & 0.7683(25)        & 0.2317                       && 0.7683(23)        & 0.2317                       \\
$L/(L_1+L_2)_{I}$               & 0.7185(26)        & 0.2815                       && 0.7185(23)        & 0.2815                       \\
$r$ (pole)                      & 0.3327(17)        & 0.3135(7)                    && 0.3343(15)        & 0.3145(7)                    \\
$r$ (point)                     & 0.3643(26)        & 0.4467(29)                   && 0.3671(24)        & 0.4481(25)                   \\
$r$ (side)                      & 0.3432(19)        & 0.3276(8)                    && 0.3451(17)        & 0.3287(7)                    \\
$r$ (back)                      & 0.3547(20)        & 0.3598(8)                    && 0.3569(20)        & 0.3609(7)                    \\
$r$ (volume)$^\dagger$          & 0.3439            & 0.3350                       && 0.3458            & 0.3361                       \\
Colatitude (deg)                &                   &                              && 71.87(85)         &                              \\
Longitude (deg)                 &                   &                              && 2.36(29)          &                              \\
Radius (deg)                    &                   &                              && 14.14(18)         &                              \\
$T$$\rm _{spot}$/$T$$\rm _{local}$  &               &                              && 1.125(9)          &                              \\
$\Sigma W(O-C)^2$               & \multicolumn{2}{c}{0.012}                        && \multicolumn{2}{c}{0.011}                       
\tablenotetext{\dagger}{Mean volume radius.}
\enddata
\end{deluxetable}

\begin{deluxetable}{cccccc}
\tablewidth{0pt}
\tablecaption{Minimum timings of CL Aur determined by two different methods.}
\tablehead{
\colhead{KvW} & \colhead{W-D$\rm$} & \colhead{Error$\rm^{a}$} & \colhead{Difference$\rm^{b}$} & \colhead{Filter} & \colhead{Min} 
}
\startdata
2,452,974.2543  &  2,452,974.25469  &  $\pm$0.00007  &  $-$0.00039  &  $BVRI$  &  I    \\
2,453,019.0518  &  2,453,019.05186  &  $\pm$0.00010  &  $-$0.00006  &  $BVRI$  &  I    \\
2,453,055.1393  &  2,453,055.13934  &  $\pm$0.00010  &  $-$0.00004  &  $BVRI$  &  I    \\
2,453,350.0577  &  2,453,350.05766  &  $\pm$0.00005  &  $+$0.00004  &  $BVRI$  &  I    \\
2,453,353.1692  &  2,453,353.16967  &  $\pm$0.00031  &  $-$0.00047  &  $BVRI$  &  II   \\
2,453,360.0128  &  2,453,360.01267  &  $\pm$0.00009  &  $+$0.00013  &  $BVRI$  &  I    \\
2,453,361.2570  &  2,453,361.25700  &  $\pm$0.00009  &  $+$0.00000  &  $BVRI$  &  I    \\
2,453,414.1438  &  2,453,414.14313  &  $\pm$0.00042  &  $+$0.00067  &  $BVRI$  &  II   \\
\enddata
\tablenotetext{a}{Errors yielded by the W-D code.}
\tablenotetext{b}{Differences between columns (1) and (2).}
\end{deluxetable}

\begin{deluxetable}{lcc}
\tablewidth{0pt} 
\tablecaption{Astrophysical parameters for CL Aur.}
\tablehead{
\colhead{Parameter}              & \colhead{Primary} & \colhead{Secondary}
}
\startdata
$M$/$M_\odot$                    &  2.24             &  1.35            \\
$R$/$R_\odot$                    &  2.58             &  2.51            \\
$\log$ $g$ (cgs)                 &  3.97             &  3.77            \\
$\log$ $\rho$/$\rho_\odot$       &  $-$0.88          &  $-$1.07         \\
$T$ (K)                          &  9,420            &  6,323           \\
$L$/$L_\odot$                    &  46.96            &  9.02            \\
$B-V$ (mag)                      &  $-$0.01          &  $+$0.48         \\
$M_{\rm bol}$ (mag)              &  $+$0.51          &  $+$2.30         \\
BC (mag)                         &  $-$0.15          &  $-$0.01         \\
$M_{V}$ (mag)                    &  $+$0.66          &  $+$2.31         \\
$M_{\rm V,total}^\dagger$ (mag)  &  \multicolumn{2}{c}{$+$0.44}         \\
Distance (pc)                    &  \multicolumn{2}{c}{1,150}           \\
\enddata
\tablenotetext{\dagger}{Absolute visual magnitude from both components.}
\end{deluxetable}

\begin{deluxetable}{lcccccc}
\tablewidth{0pt} 
\tablecaption{Impact position and energy of transferred matter.}
\tablehead{
\multicolumn{3}{c}{Initial velocity (km s$^{-1}$)}   && \multicolumn{2}{c}{Impact position (deg)}   & \colhead{Impact energy ($L_\odot$)}    \\ [1.0mm] \cline{1-3} \cline{5-6} \\[-2.0ex]
\colhead{$v_x$} & \colhead{$v_y$} & \colhead{$v_z$}  && \colhead{Longitude} & \colhead{Colatitude}  & 
}
\startdata
3$\times 10^{-7}$  &  0  &  0                   && 10.5  &  90.0  &  0.69     \\
3$\times 10^{-5}$  &  0  &  0                   && 10.3  &  90.0  &  0.69     \\
3$\times 10^{-3}$  &  0  &  0                   && 10.5  &  90.0  &  0.69     \\
3$\times 10^{-1}$  &  0  &  0                   && 10.6  &  90.0  &  0.70     \\
3                  &  0  &  0                   && 10.5  &  90.0  &  0.71     \\
30                 &  0  &  0                   && 11.3  &  90.0  &  0.73     \\
300                &  0  &  0                   && 4.9   &  90.0  &  1.85     \\ 
450                &  0  &  0                   && 3.4   &  90.0  &  3.24     \\ 
3$\times 10^{-1}$  &  0  &  3$\times 10^{-1}$   && 10.6  &  90.0  &  0.70     \\ 
3                  &  0  &  3                   && 10.6  &  90.3  &  0.71     \\                  
30                 &  0  &  30                  && 11.4  &  85.3  &  0.72     \\                  
150                &  0  &  150                 && 12.4  &  65.4  &  1.26     \\                  
300                &  0  &  300                 && 20.6  &  43.6  &  2.96     \\                  
\enddata
\end{deluxetable}


\newpage
\begin{thebibliography}{}
\bibitem[Agerer \& Hubscher(2002)]{agerer2002} Agerer, F., \& H\"ubscher, J. 2002, Inf. Bull. Variable Stars, No. 5296
\bibitem[Applegate(1992)]{applegate1999} Applegate, J. H. 1992, ApJ, 385, 621
\bibitem[Baldwin \& Samolyk(2004)]{baldwin2004} Baldwin, M. E., \& Samolyk, G. 2004, Observed Minima Timings of Eclipsing Binaries, No. 9 (Cambridge: AAVSO)
\bibitem[Baldwin \& Samolyk(2007)]{baldwin2007} Baldwin, M. E., \& Samolyk, G. 2007, Observed Minima Timings of Eclipsing Binaries, No. 12 (Cambridge: AAVSO)
\bibitem[B\'ir\'o et al(2007)]{biro2007} B\'ir\'o, I. B., et al. 2007, Inf. Bull. Variable Stars, No. 5753
\bibitem[Blattler(2000)]{blattler2000} Bl\"attler, E. 2000, BBSAG Bull., No. 121
\bibitem[Borkovits(2008)]{borkovits2008} Borkovits, T., et al. 2008, Inf. Bull. Variable Stars, No. 5835
\bibitem[Brat et al(2007)]{brat2007} Br\'at, L., Zejda, M., \& Svoboda, P. 2007, Open Eur. J. Var. Stars, 74, 1
\bibitem[Brat et al(2008)]{brat2008} Br\'at, L., et al. 2008, Open Eur. J. Var. Stars, 94, 1
\bibitem[Brat et al(2009)]{brat2009} Br\'at, L., et al. 2009, Open Eur. J. Var. Stars, 107, 1
\bibitem[De Loore \& van Rensbergen(2005)]{de2005} De Loore, C., \& van Rensbergen, W. 2005, Ap\&SS, 296, 353
\bibitem[Diethelm(2001)]{diethelm2001} Diethelm, R. 2001, BBSAG Bull., No. 126
\bibitem[Dogru et al(2007)]{dogru2007} Do\u{g}ru, S. S., D\"onmez, A., T\"uys\"uz, M., Do\u{g}ru, D., \"Ozkarde\c s, B., Soydugan, E., \& Soydugan, F. 2007, Inf. Bull. Variable Stars, No. 5746
\bibitem[ESA(1997)]{esa1997} ESA. 1997, The Hipparcos and Tycho Catalogues (ESA SP-1200; Noordwijk: ESA)
\bibitem[Girardi et al(2000)]{girardi2000} Girardi, L., Bressan, A., Bertelli, G., \& Chiosi, C. 2000, A\&AS, 141, 371
\bibitem[G\"otz \& Wenzel(1968)]{gotz1968} G\"otz, W., \& Wenzel, W. 1968, Mitteilungen Ver\"anderliche Sterne, 5, 5
\bibitem[Hall(1989)]{hall1989} Hall, D. S. 1989, Space Sci. Rev., 50, 219
\bibitem[Harmanec(1988)]{harmanec1988} Harmanec, P. 1988, Bull. Astron. Inst. Czechoslovakia, 39, 329
\bibitem[Hegedus(1988)]{hegedus1988} Heged\"us, T. 1988, Bull. Inf. CDS, 35, 15
\bibitem[Hilditch(1989)]{hilditch1989} Hilditch, R. W. 1989, Space Sci. Rev., 50, 289
\bibitem[Hoffleit(1935)]{hoffleit1935} Hoffleit, D. 1935, Harvard Bulletin, 901, 20
\bibitem[Hog(2000)]{hog2000} H\o g, E., et al. 2000, A\&A, 355, 27
\bibitem[H\"ubscher(2007)]{hubscher2007a} H\"ubscher, J. 2007, Inf. Bull. Variable Stars, No. 5802
\bibitem[H\"ubscher et al(2005)]{hubscher2005} H\"ubscher, J., Paschke, A., \& Walter, F. 2005, Inf. Bull. Variable Stars, No. 5657
\bibitem[H\"ubscher et al(2006)]{hubscher2006} H\"ubscher, J., Paschke, A., \& Walter, F. 2006, Inf. Bull. Variable Stars, No. 5731
\bibitem[H\"ubscher \& Walter(2007)]{hubscher2007b} H\"ubscher, J., \& Walter, F. 2007, Inf. Bull. Variable Stars, No. 5761
\bibitem[Irwin(1952)]{irwin1952} Irwin, J. B. 1952, ApJ, 116, 211
\bibitem[Irwin(1959)]{irwin1959} Irwin, J. B. 1959, AJ, 64, 149
\bibitem[Kang et al(2007)]{kang2007} Kang, Y.-W., Hong, K.-S., \& Lee, J. 2007, in ASP Conf. Ser. 362, The Seventh Pacific Rim Conference on Stellar Astrophysics, ed. Y.-W. Kang et al. (San Francisco: ASP), 19
\bibitem[Kim et al(1997)]{kim1997} Kim, C.-H., Jeong, J. H., Demircan, O., M\"{u}yessero\u{g}lu, Z., \& Budding, E. 1997, AJ, 114, 2753
\bibitem[Kim et al(2006)]{kim2006} Kim, C.-H., Lee, C.-U., Yoon, Y.-N., Park, S.-S., Kim, D.-H., Cha, S.-M., \& Won, J.-H. 2006, Inf. Bull. Variable Stars, No. 5694
\bibitem[Kim et al(2003)]{kim2003} Kim, S.-L., Lee, J. W., Kwon, S.-G., Youn, J.-H., Mkrtichian, D. E., Kim, C. 2003, A\&A, 405, 231
\bibitem[Kim et al(2002)]{kim2002} Kim, K.-M. et al., 2002, J. Korea Astron. Soc., 35, 221
\bibitem[Kreiner et al(2001)]{kreiner2001} Kreiner, J. M., Kim, C.-H., \& Nha, I.-S. 2001, An Atlas of $O$--$C$ Diagrams of Eclipsing Binary Stars (Krakow: Wydawn. Nauk. Akad. Pedagogicznej)
\bibitem[Kurochkin(1951)]{kurochkin1951} Kurochkin, N. E. 1951, Variable Stars, 8, 351
\bibitem[Kwee \& van Woerden(1956)]{kwee1956} Kwee, K. K., \& van Woerden, H. 1956, Bull. Astron. Inst. Netherlands, 12, 327
\bibitem[Lanza, Rodono \& Rosner(1998)]{lanza1998} Lanza, A. F., Rodono, M., \& Rosner, R. 1998, MNRAS, 296, 893
\bibitem[Lee et al(2007)]{lee2007} Lee, J. W., Kim, C.-H., \& Koch, R. H. 2007, MNRAS, 379, 1665
\bibitem[Lee et al(2009a)]{lee2009a} Lee, J. W., Kim, S.-L., Lee, C.-U., Kim, H.-I., Park, J.-H., Park, S.-R., \& Koch, R. H. 2009a, PASP, 121, 104
\bibitem[Lee et al(2009b)]{lee2009b} Lee, J. W., Youn, J.-H., Lee, C.-U., Kim, S.-L., \& Koch, R. H. 2009b, AJ, 138, 478
\bibitem[Lenz \& Breger(2005)]{lenz2005} Lenz, P., \& Breger, M. 2005, Comm. Asteroseismology, 146, 53
\bibitem[Lubow \& Shu(1975)]{lubow1975} Lubow, S. H., \& Shu, F. H. 1975, ApJ, 198, 383
\bibitem[Maceroni \& van't Veer(1994)]{maceroni1994} Maceroni, C., \& van't Veer, F. 1994, A\&A, 289, 871
\bibitem[Mkrtichian(2004)]{mkrtichian2004} Mkrtichian, D. E., et al. 2004, A\&A, 419, 1015
\bibitem[Mochnacki(1984)]{mochnacki1984} Mochnacki, S. W. 1984, ApJS, 55, 551
\bibitem[Nelson(2006)]{nelson2006} Nelson, R. H. 2006, Inf. Bull. Variable Stars, No. 5672
\bibitem[Press et al(1992)]{press1992} Press, W. H., Teukolsky, S. A., Vetterling, W. T., \& Flannery, B. P. 1992, Numerical Recipes (Cambridge: Cambridge Univ. Press), Chapter 15
\bibitem[Richards \& Albright(1999)]{richards1999} Richards, M. T., \& Albright, G. E. 1999, ApJS, 123, 537
\bibitem[Rodriguez et al(2004)]{rodriguez2004} Rodriguez, E., et al. 2004, MNRAS, 347, 1317
\bibitem[Smith \& Caton(2007)]{smith2007} Smith, A. B., \& Caton, D. B. 2007, Inf. Bull. Variable Stars, No. 5745
\bibitem[Soydugan et al(2006)]{soydugan2006} Soydugan, E., Soydugan, F., Demircan, O., \& \. Ibanoglu, C. 2006, MNRAS, 370, 2013
\bibitem[Van Hamme(1993)]{van1993} Van Hamme, W. 1993, AJ, 106, 209
\bibitem[Wilson \& Biermann(1976)]{wilson1976} Wilson, R. E., \& Biermann, P. 1976, A\&A, 48, 349
\bibitem[Wilson \& Devinney(1971)]{wilson1971} Wilson, R. E., \& Devinney, E. J. 1971, ApJ, 166, 605
\bibitem[Wolf et al(1999)]{wolf1999} Wolf, M., \u{S}arounov\'a, L., Bro\u{z}, M., \& Horan, R. 1999, Inf. Bull. Variable Stars, No. 4683
\bibitem[Wolf et al(2007)]{wolf2007} Wolf, M., et al. 2007, Inf. Bull. Variable Stars, No. 5780 (W07)
\end{thebibliography}
\end{document}